\documentclass[aps,pre,preprint,groupedaddress]{revtex4-1}

\usepackage{graphicx}
\usepackage{float} 
\usepackage{setspace} 
\usepackage{amssymb} 
\usepackage{amsmath} 
\usepackage[hidelinks]{hyperref}

\begin{document}

\noindent{\bf\Large A small PAM optimises target recognition in the \\ CRISPR-Cas immune system} \\

\noindent {\large Melia E.\ Bonomo$^{1,2}$} \\

\noindent$^{1}$Department of Physics \& Astronomy, Rice University, Houston, TX, USA \\
$^{2}$Center for Theoretical Biological Physics, Rice University, Houston, TX, USA \\


\noindent{\bf Author for correspondence:} M.E.\ Bonomo, mbonomo@rice.edu

\section*{Abstract}

\noindent CRISPR-Cas is an adaptive immune mechanism that has been harnessed for a variety of genetic engineering applications: the Cas9 protein recognises a 2--5nt DNA motif, known as the PAM, and a programmable crRNA binds a target DNA sequence that is then cleaved. While off-target activity is undesirable, it occurs because cross-reactivity was beneficial in the immune system on which the machinery is based. Here, a stochastic model of the target recognition reaction was derived to study the specificity of the innate immune mechanism in bacteria. CRISPR systems with Cas9 proteins that recognised PAMs of varying lengths were tested on self and phage DNA. The model showed that the energy associated with PAM binding impacted mismatch tolerance, cleavage probability, and cleavage time. Small PAMs allowed the CRISPR to balance catching mutant phages, avoiding self-targeting, and quickly dissociating from critically non-matching sequences. Additionally, the results revealed a lower tolerance to mismatches in the PAM and a PAM-proximal region known as the seed, as seen in experiments. This work illustrates the role that the Cas9 protein has in dictating the specificity of DNA cleavage that can aid in preventing off-target activity in biotechnology applications.

\section{Introduction}

Clustered regularly interspaced short palindromic repeats (CRISPR) and the CRISPR-associated proteins (Cas) constitute a genetic adaptive immune system found in bacteria and archaea~\cite{makarova2020,bonomo2018}.  The immunological memory is comprised of alternating DNA repeats and short sequences known as spacers, which match sequences known as protospacers in the genomes of mobile genetic threats, such as phages.  CRISPR-Cas transcribes its spacers as RNA guides (crRNA) that help Cas proteins to recognise and cleave subsequent genetic infectors.  Owing to its ability to interact with programmable DNA targets, the Cas9 protein from \emph{Streptococcus pyogenes} has become a prominent tool for a variety of genetic editing and gene expression applications~\cite{doudna2014}.  As these biotechnologies rapidly advance, there is still more to be understood about the specificity and efficiency of DNA recognition and cleavage.

The CRISPR target interrogation reaction is modular: the Cas9 searches for and binds to a 2--5nt protospacer-associated DNA motif (PAM), and the crRNA binds to the associated DNA sequence~\cite{jones2017}. During the PAM interaction, the Cas9 domain that locks with the phosphate group of the first DNA base pair causes a distortion in the double-strands that starts double-stranded DNA (dsDNA) melting~\cite{anders2014}.  Melting relies on Cas9:PAM binding energy, because Cas9 does not hydrolyse ATP~\cite{sternberg2014}.  After the start of local dsDNA melting, crRNA:DNA hybridisation begins, where the roughly 30nt crRNA binds the available target DNA in a directional, base-by-base sequential manner~\cite{sternberg2014,farasat2016}. A sufficient crRNA:DNA match will lead to cleavage of the target DNA. Experiments and stochastic modelling have shown that the number and position of mismatches between the Cas9:crRNA and target DNA impact the CRISPR's specificity, whereby multiple mismatches in the PAM and PAM-proximal region significantly lower cleavage probability~\cite{jiang2013,farasat2016,klein2018}. 

To date, there has been little work investigating the extent to which the Cas9:crRNA modularity regulates a successful immune recognition reaction. Here, a stochastic model was derived to study how the size of the PAM module impacts the free energy landscape of the target interrogation reaction, which was defined as beginning with Cas9:PAM binding, followed by base-by-base dsDNA melting and crRNA:DNA binding, and ending with DNA cleavage. Cleavage probability, cleavage time, and dissociation time (when cleavage does not occur) were calculated for simulated phage DNA with varying mutation rates and self DNA. The model demonstrated that small PAMs, comparable to those found in endogenous CRISPR systems, exhibited hierarchical mismatch tolerance and were sufficiently specific for determining self versus non-self DNA, while still being sufficiently cross-reactive and fast to protect against mutant phage attacks.

\section{Methods}

\subsection{The DNA Target Recognition Reaction}

\begin{figure}[!h]
\begin{center}
\includegraphics[width=\textwidth]{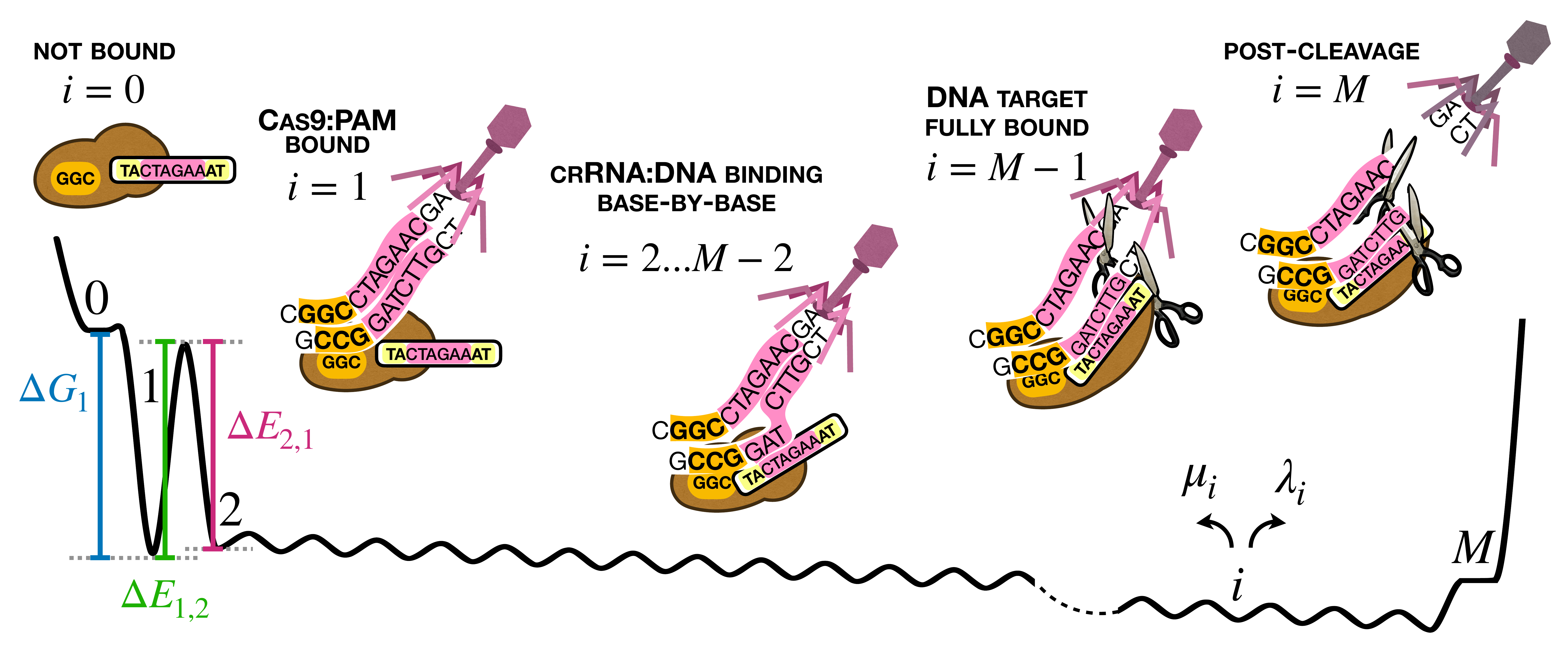}
\caption{Diagram of the reaction landscape for CRISPR-Cas interrogation of a DNA sequence. State 0 represents the Cas9:crRNA unbound to the DNA. In state 1, the Cas9 has bound the PAM, and in states $i=2$ through $M-2$, there is binding of individual crRNA:DNA base pairs. State $M-1$ represents completed crRNA:DNA hybridisation, and state $M$ is after cleavage occurs. Examples of the free energy $\Delta{}G_{i-1}$ and barrier heights $\Delta{}E_{i-1,i}$ and $\Delta{}E_{i,i-1}$ for $i=2$ are marked.}
\label{fig:landscape}
\end{center}
\end{figure}

Let us consider the reaction through which the CRISPR-Cas binds and cleaves a DNA sequence as unfolding on a landscape of 0 to $M$ discrete states (Figure~\ref{fig:landscape}). The barrier heights $\Delta{}E_{j,k}$ between state $j$ to state $k$ are determined as
\begin{align}
\Delta{}E_{i,i-1}&= \Delta{}E_{i-1,i} - \Delta{}G_i 
\label{eq:DeltaEbackward}
\end{align}
where $\Delta{}G_i$ is the free energy difference between state $i$ and $i-1$, defined by
\begin{align}
\Delta{}G_1 &= \Delta{}G_{\textrm{PAM}}(n_{\text{match}}), \\
\Delta{}G_{1 > i > M} &= \Delta{}G_{\textrm{dsDNAseparation}} + \Delta{}G_{\textrm{BindDNA}}, \\
\Delta{}G_{M} &= \Delta{}G_{\textrm{Cleavage}}.
\label{eq:DeltaG}
\end{align}
The $\Delta{}G_{\textrm{PAM}}(n_{\text{match}})$ is a the free energy associated with binding the Cas9:DNA, where $n_{\text{match}}$ is the number of matches between the Cas9 and the target DNA sequence's PAM. The free energy of melting each target DNA base pair, $\Delta{}G_{\textrm{dsDNAseparation}}$, is based on the energy associated with melting dsDNA that is matched and negatively supercoiled, which makes it energetically favourable to separate the strands~\cite{farasat2016}.  The free energy $\Delta{}G_{\textrm{BindDNA}}$ associated with binding each crRNA:DNA base pair after the PAM is $\Delta{}G_{\textrm{BindMatch}}$ if the base pair matches and $\Delta{}G_{\textrm{BindMismatch}}$ if it does not.  The $\Delta{}G_{\textrm{Cleavage}}$ is the free energy associated with breaking the phosphodiester bonds of both strands of the DNA target. The reverse $\mu$ and forward $\lambda$ rates for each reaction state are then determined using the Arrhenius relation as
\begin{align}
\mu_i &= A_{i,i-1} e^{-\Delta{}E_{i,i-1}/k_B T} \\
\lambda_i &= A_{i,i+1} e^{-\Delta{}E_{i,i+1}/k_B T},
\end{align}
where $A_{j,k}$ is the attempt rate to cross the barrier from state $j$ to state $k$, $k_B$ is the Boltzmann constant, and $T$ is temperature.  
The free energy and kinetic parameters used in the model are estimated from experiments, as described in the Supplemental Material.

\subsection{Probability and Time to Cleave DNA Target}

Utilising the reverse and forward reaction rates, the probability of cleaving a DNA target from the initial Cas9:PAM interaction in state $i=1$ is
\begin{equation}
p_{(1,M)} = \frac{1}{1 + \sum\limits^{M-1}_{i=1} \prod\limits^{i}_{j=1} \gamma_j},
\label{eq:pCleave}
\end{equation}
and from any state $i$ is
\begin{equation}
p_{(i,M)} = p_{(1,M)} \Big( 1 + \sum^{i-1}_{j=1} \prod^{j}_{k=1} \gamma_k \Big),
\end{equation}
where $\gamma_i = \mu_i / \lambda_i$.  The time to cleave the DNA target starting from the initial Cas9:PAM state $i=1$ is
\begin{equation}
t_{(1,M)} = \sum\limits^{M-1}_{i=1}\frac{p_{(i,M)}}{\lambda_i} + \sum\limits^{M-2}_{i=1}\frac{p_{(i,M)}}{\lambda_i} \Bigg( \sum\limits^{M-1}_{j=i+1} \prod\limits^{j}_{k=i+1} \gamma_k  \Bigg),
\label{eq:tCleave}
\end{equation}
and the time for the Cas9 protein to dissociate from the PAM is
\begin{equation}
t_{(1,0)} = \Bigg[ \sum\limits^{M-1}_{i=1}\frac{1-p_{(i,M)}}{\lambda_i} + \sum\limits^{M-2}_{i=1}\frac{1-p_{(i,M)}}{\lambda_i} \Bigg( \sum\limits^{M-1}_{j=i+1} \prod\limits^{j}_{k=i+1} \gamma_k  \Bigg) \Bigg] \frac{p_{(1,M)}}{1-p_{(1,M)}}.
\label{eq:tDissociate}
\end{equation}
These equations were derived using the backwards Fokker-Planck equation, as detailed in the Supplemental Material, and they describe the absorbing probability and time for any stochastic system with state-dependent forward and reverse rates.

\section{Modelling Results} 

Hypothetical Cas9 proteins that had PAM specificities varying in length from 0nt to 33nt were associated with crRNAs and tested on target DNA sequences of fixed length, consisting of 33nt for the PAM and protospacer together, with assorted numbers and locations of mismatches. For the Cas9:crRNA interrogation of each DNA sequence, the landscape of reaction states was generated (Supplemental Figure S1), the probability of cleavage was calculated with Eq.~\ref{eq:pCleave}, and a random number was generated to determine whether or not cleavage occurred given this probability. If the target sequence was cleaved, the cleavage time was calculated with Eq.~\ref{eq:tCleave}, otherwise the time for dissociation of the Cas9:crRNA from the target DNA sequence was calculated with Eq.~\ref{eq:tDissociate}.

\subsection{Random Mismatches}

In the first run of the model, mismatches were generated in random sequence locations. Phage PAMs and protospacers were given mutation rates $\nu=0.05, 0.15, 0.30,$ and 0.50, corresponding to approximately 2, 5, 10, and 16 mismatches in each sequence, respectively. Self DNA sequences of the same 33nt-length were generated with arbitrary matches to the Cas9:crRNA with a 1 in 4 probability (equivalent to $\nu=0.75$), corresponding to 24 mismatches on average.

\begin{figure}[h!]
\begin{center}
{\bf A}\includegraphics[width=0.45\textwidth]{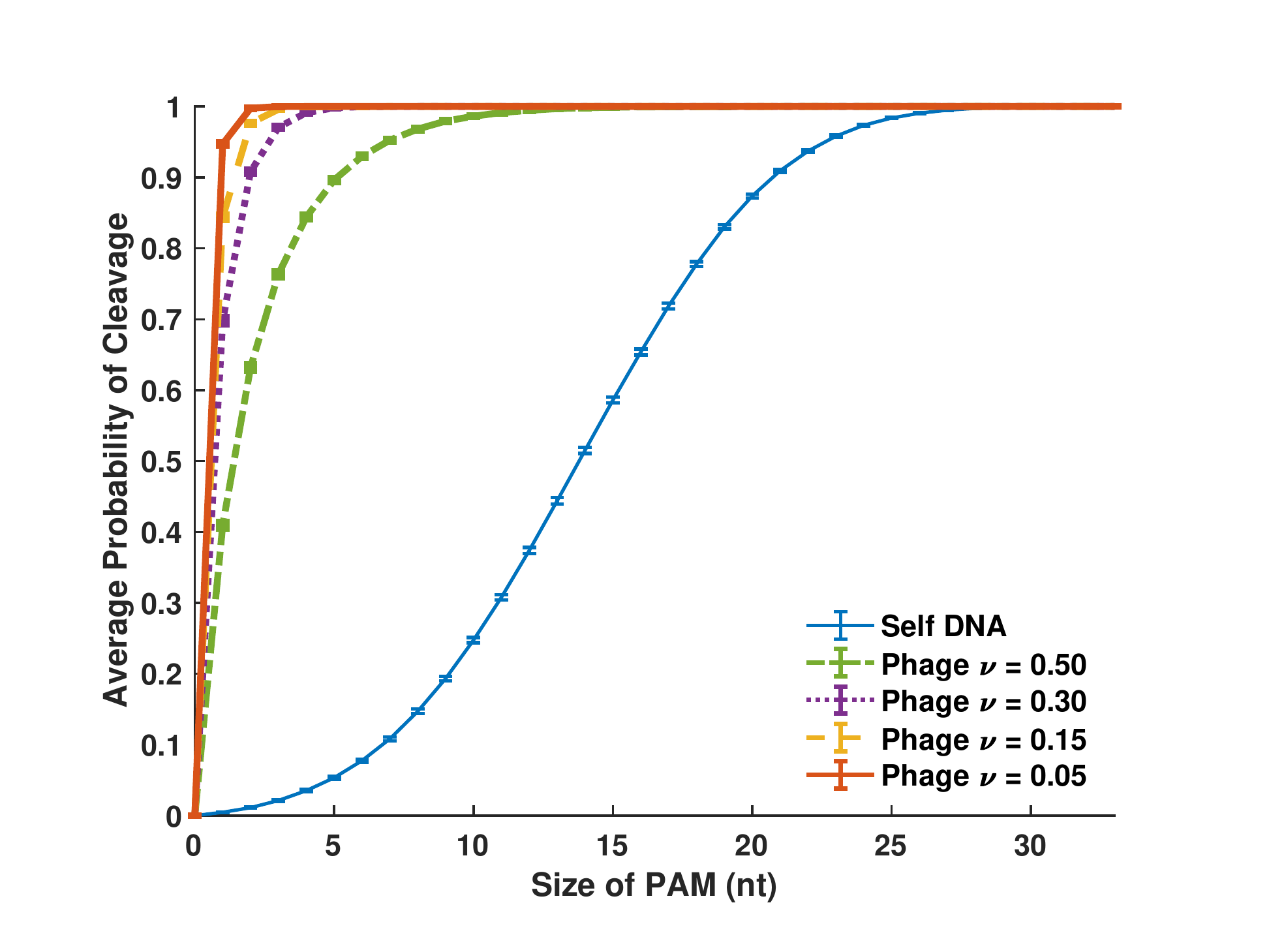}
{\bf B}\includegraphics[width=0.45\textwidth]{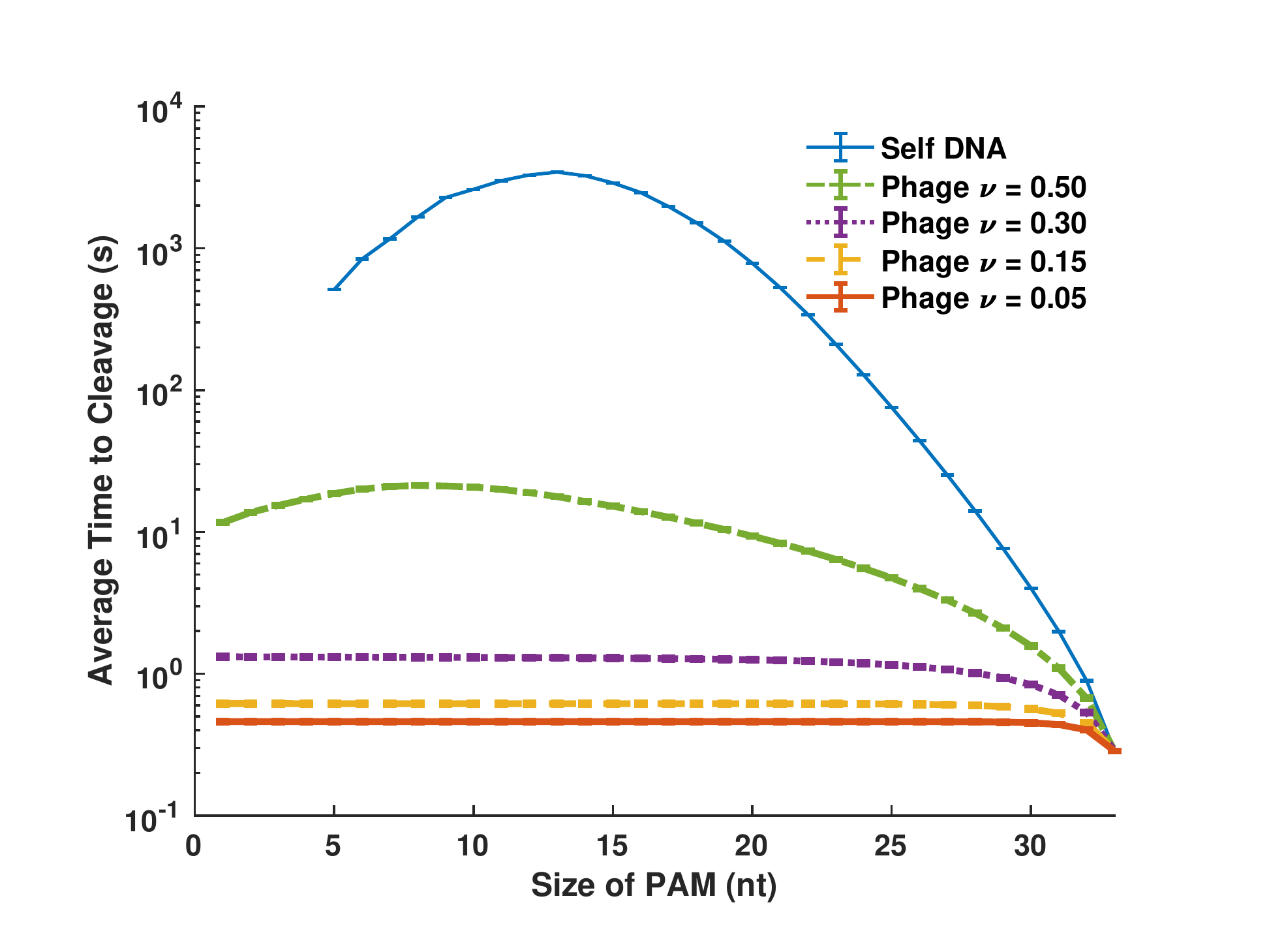}
{\bf C}\includegraphics[width=0.45\textwidth]{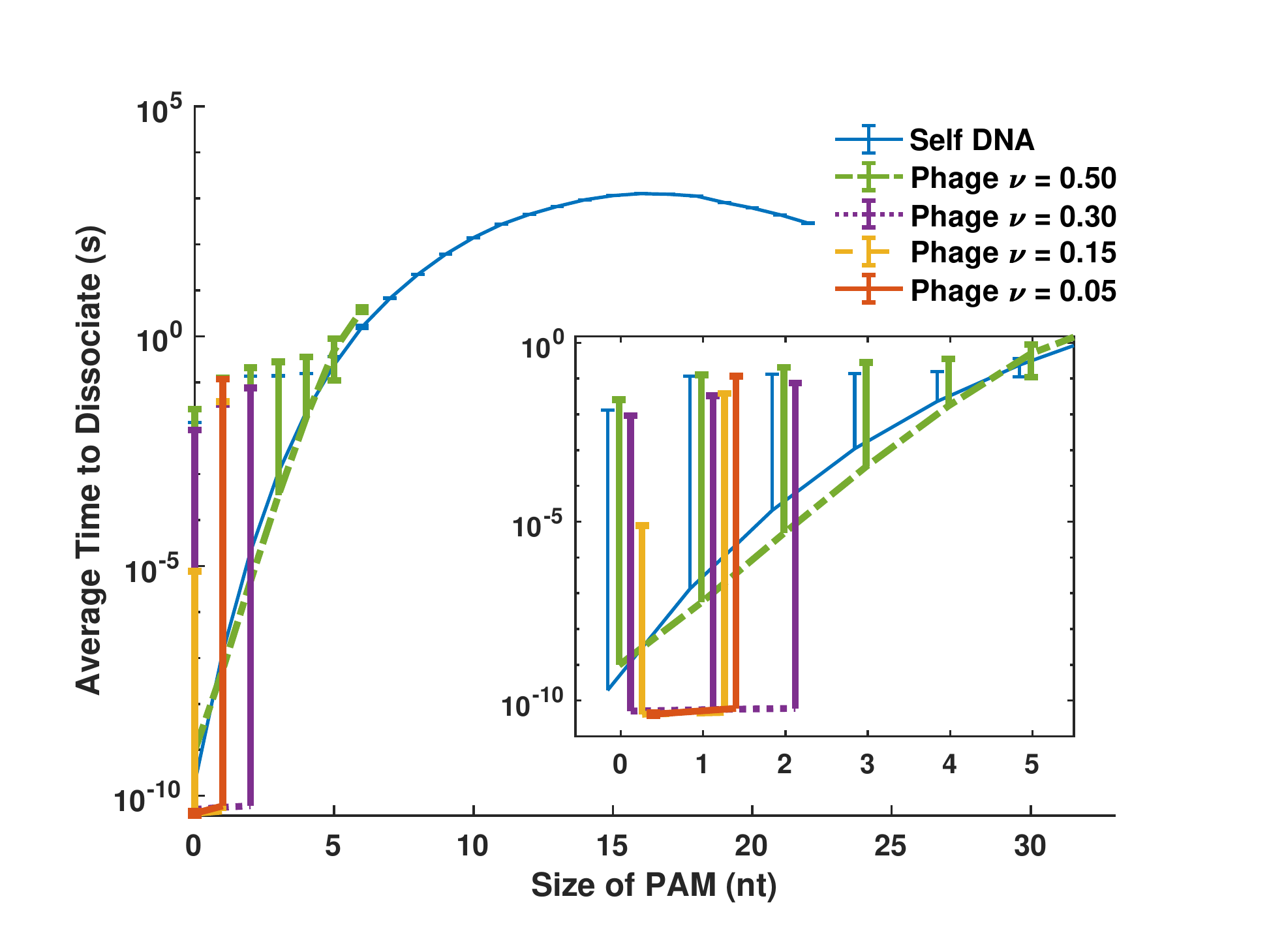}
\caption{{\bf (A)} Probability of cleavage, {\bf (B)} cleavage time, and {\bf (C)} dissociation time when cleavage did not occur (inset shows the curves offset for clarity). Each curve was averaged from 10,000 iterations of the model, and error bars represent standard error. The average times were calculated as geometric means when the respective events occurred in at least 5\% of the iterations for a particular PAM size.}
\label{fig:pCleave-tCleave}
\end{center}
\end{figure}

PAMs that were at least 2nt had $p_{(1,M)}>0.5$ for phage targets at all tested mutation rates (Fig.~\ref{fig:pCleave-tCleave}A). Large PAMs that made up over half of the fixed length DNA target were not deterred from cleaving any of the phage DNA, and those that were greater than 25nt cleaved every target, regardless of the number of mismatches. The cleavage time of phage DNA with $\nu \le 0.30$ was not greatly affected by PAM size, though in general, more mismatches slowed down cleavage (Fig.~\ref{fig:pCleave-tCleave}B). Importantly, the PAMs that were 5nt or less had $p_{(1,M)}<0.05$ for cleaving their own sequences, and a very fast dissociation time (on the order of $10^{-10}$s to $10^{-1}$s) from any sequence that was not cleaved (Fig.~\ref{fig:pCleave-tCleave}C).

\subsection{Consecutive Mismatches}

\begin{figure}[h!]
\begin{center}
\includegraphics[width=\textwidth]{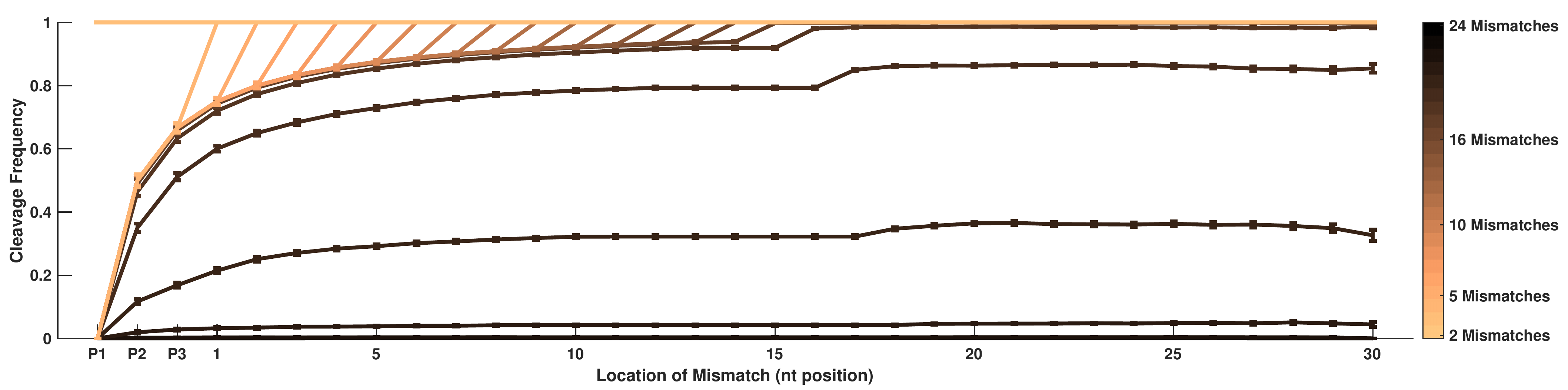}
\caption{Mismatch tolerance for the 3-nt PAM system illustrated by averaging over cleavage occurrences when there were mismatches at each sequence location. A cleavage frequency of 1 means that a mismatch at that particular location was always tolerated, whereas 0 means it was never tolerated, when there were the specified number of consecutive mismatches. Each curve was averaged from 10,000 iterations of the model, and error bars represent standard error. P$i$ denotes the location of PAM nucleotide $i$, and the $x$-axis restarts for nucleotides after the PAM.}
\label{fig:mismatches}
\end{center}
\end{figure}

In the second model run, mismatches were generated consecutively across all possible positions. Mismatch tolerance was profiled for all numbers of mismatches from 2 to 24, by averaging over the number of times cleavage occurred (called cleavage frequency) when there was a mismatch at a particular sequence location. For the 3nt-PAM system, there was low tolerance (cleavage frequency $< 0.6$) in the PAM and moderate tolerance (cleavage frequency $\approx 0.8$) in the first 15nt after the PAM depending on the number of consecutive mismatches (Fig.~\ref{fig:mismatches}).

\section{Discussion}

An effective CRISPR system needs to be able to swiftly cleave foreign DNA, including sequences that are not exact matches to the crRNA, while avoiding cleaving self DNA and quickly dissociating to minimise the DNA target search time. This work showed how the modularity of Cas9:PAM and crRNA:DNA binding regulated target recognition from the energy associated with binding the first module.

Larger PAMs led to more cross-reactivity, whereas smaller PAMs led to more specific targeting. Experiments have shown that Cas9 evolved with broadened PAM compatibility led to higher DNA targeting specificity~\cite{hu2018}. In other words, the engineered Cas9 had a less specific PAM interaction, resulting in lower binding energy than the wild type, and engaged in less off-target activity as a result. Furthermore, the heightened sensitivity to mismatches in the PAM and PAM-proximal region has been experimentally observed. Single mismatches in the PAM increase the binding free energy but can still lead to crRNA:DNA hybridisation~\cite{farasat2016,sternberg2014}. In particular, if there is sufficient crRNA:DNA complementarity in the first 8--12 bp, known as the seed region, and up to 8 mismatches in the remainder of the sequence~\cite{sternberg2014}.  

CRISPR systems with smaller PAMs were just as fast as those with larger PAMs at cleaving phage DNA at low mutation rates, and they had fast dissociation times, though with large variation. \emph{In vivo} experiments with a single Cas9 in \emph{Escherichia coli} demonstrated that interrogating a sequence next to a PAM site took less than 30 ms on average~\cite{jones2017}. However, while some Cas9:crRNA complexes dissociate immediately when the seed does not match, others engage in more of crRNA:DNA hybridisation before dissociating~\cite{szczelkun2014}. Additionally, while PAM-distal mismatches did not affect cleavage probability, the model demonstrated that $\ge$2 mismatches at the end of the sequence increased the cleavage time several orders of magnitude (Supplemental Figure S2). This is broadly in line with experimental observations that 1--9-bp truncations off the end of the crRNA reduced dsDNA cleavage rates~\cite{szczelkun2014}.

There are several limitations to this model. First, beyond regulating the tradeoff between autoimmunity and cross-reactivity, the size of the PAM likely has other constraints. For instance, just as there is a cost associated with maintaining a sufficient repertoire of CRISPR spacers~\cite{bradde2020}, there would be a significant cost associated with maintaining Cas9 proteins with longer, more specific PAM targets. Second, the model does not currently take the Cas9 search mechanism into account. Theoretical modelling has suggested there is an optimal balance of interactions, whereby binding a longer PAM would lead to a stronger and more specific initial reaction, however the search process for finding a matching target would be slowed down~\cite{shvets2017}. The results here do show that Cas9 with longer PAMs take more time to dissociate from sequences that have many mismatches, and thus including the full search process would refine our understanding of the first module and overall CRISPR efficiency. Third, the model currently accounts for a simple Cas9:PAM interaction, however a more detailed approach would include protein morphology dynamics and the additional interactions observed in experiments~\cite{nishimasu2015}.  Topological distortion of the protein when there are mismatches at particular sequence locations affects the associated energy cost~\cite{jiang2015}. Finally, experimental observations suggest there is a second reaction pathway during Cas9:crRNA interrogation of a DNA target: primed acquisition, which is when the CRISPR recognises it has bound an imperfect match and more spacers are collected rather than cleavage or dissociation occurring~\cite{fineran2012}. The model presented here could be used to study the mismatch profile that leads to primed acquisition.

In conclusion, stochastic modelling of the modular CRISPR target recognition reaction demonstrated that Cas9 proteins that recognise 1--5nt PAMs are sufficiently cross-reactive to catch escape mutants while largely avoiding self-targeting; they also quickly dissociate from non-matching sequences. Hierarchical mismatch tolerance emerged, with the highest sensitivity to PAM mismatches, followed by those in a $\approx$15nt seed region. While cross-reactivity is beneficial in an immune system, it can cause undesirable off-target activity when the machinery is used for genetic engineering applications. The model suggests that the energy associated with the PAM interaction is instrumental in managing this, as the smaller the PAM, the lower the binding energy and the less off-target activity.

\vskip6pt

\enlargethispage{20pt}

{\bf Competing Interests.} The author declares that the research was conducted in the absence of any commercial or financial relationships that could be construed as a potential conflict of interest.

{\bf Funding.} No funding was received for this study.

{\bf Acknowledgements.} The author would like to thank M.W.\ Deem for valuable guidance in developing the model and A.B.\ Kolomeisky for helpful discussions. Part of this work appears in the author's doctoral thesis~\cite{bonomoPhD}.

\vspace*{-5pt}


\bibliographystyle{unsrtnatabbv}
\bibliography{Bonomo_CRISPR}{}

\begin{thebibliography}{17}
\providecommand{\natexlab}[1]{#1}
\providecommand{\url}[1]{\texttt{#1}}
\expandafter\ifx\csname urlstyle\endcsname\relax
  \providecommand{\doi}[1]{doi: #1}\else
  \providecommand{\doi}{doi: \begingroup \urlstyle{rm}\Url}\fi

\bibitem[Makarova et~al.(2020)Makarova, Wolf, Iranzo, Shmakov, Alkhnbashi,
  Brouns, Charpentier, Cheng, Haft, Horvath, and et]{makarova2020}
Makarova KS, Wolf YI, Iranzo J, Shmakov SA, Alkhnbashi OS, Brouns SJJ,
  Charpentier E, Cheng D, Haft DH, Horvath P, et~al.
\newblock {Evolutionary classification of CRISPR--Cas systems: a burst of class
  2 and derived variants}.
\newblock \emph{Nat Rev Microbiol}, 18\penalty0 (2):\penalty0 67--83, 2020.

\bibitem[Bonomo and Deem(2018)]{bonomo2018}
Bonomo ME, Deem MW.
\newblock {The physicist's guide to one of biotechnology's hottest new topics:
  CRISPR-Cas}.
\newblock \emph{{Phys Biol}}, 15\penalty0 (4), 2018.

\bibitem[Doudna and Charpentier(2014)]{doudna2014}
Doudna JA, Charpentier E.
\newblock {The new frontier of genome engineering with CRISPR-Cas9}.
\newblock \emph{Science}, 346\penalty0 (6213), 2014.

\bibitem[Jones et~al.(2017)Jones, Leroy, Unoson, Fange, {\'C}uri{\'c}, Lawson,
  and Elf]{jones2017}
Jones DL, Leroy P, Unoson C, Fange D, {\'C}uri{\'c} V, Lawson MJ, Elf J.
\newblock {Kinetics of dCas9 target search in \emph{Escherichia} coli}.
\newblock \emph{{Science}}, 357\penalty0 (6358):\penalty0 1420--1424, 2017.

\bibitem[Anders et~al.(2014)Anders, Niewoehner, Duerst, and Jinek]{anders2014}
Anders C, Niewoehner O, Duerst A, Jinek M.
\newblock {Structural basis of PAM-dependent target DNA recognition by the Cas9
  endonuclease}.
\newblock \emph{Nature}, 513\penalty0 (7519):\penalty0 569--573, 2014.

\bibitem[Sternberg et~al.(2014)Sternberg, Redding, Jinek, Greene, and
  Doudna]{sternberg2014}
Sternberg SH, Redding S, Jinek M, Greene EC, Doudna JA.
\newblock {DNA interrogation by the CRISPR RNA-guided endonuclease Cas9}.
\newblock \emph{{Nature}}, 507\penalty0 (7490):\penalty0 62, 2014.

\bibitem[Farasat and Salis(2016)]{farasat2016}
Farasat I, Salis HM.
\newblock {A biophysical model of CRISPR/Cas9 activity for rational design of
  genome editing and gene regulation}.
\newblock \emph{PLoS Comp Biol}, 12\penalty0 (1):\penalty0 e1004724, 2016.

\bibitem[Jiang et~al.(2013)Jiang, Bikard, Cox, Zhang, and
  Marraffini]{jiang2013}
Jiang W, Bikard D, Cox D, Zhang F, Marraffini LA.
\newblock {RNA-guided editing of bacterial genomes using CRISPR-Cas systems}.
\newblock \emph{Nat Biotechnol}, 31\penalty0 (3):\penalty0 233--239, 2013.

\bibitem[Klein et~al.(2018)Klein, Eslami-Mossallam, Arroyo, and
  Depken]{klein2018}
Klein M, Eslami-Mossallam B, Arroyo DG, Depken M.
\newblock {Hybridization Kinetics Explains CRISPR-Cas Off-Targeting Rules}.
\newblock \emph{{Cell Rep}}, 22\penalty0 (6):\penalty0 1413--1423, 2018.

\bibitem[Hu et~al.(2018)Hu, Miller, Geurts, Tang, Chen, Sun, Zeina, Gao, Rees,
  Lin, and et]{hu2018}
Hu~JH, Miller SM, Geurts MH, Tang W, Chen L, Sun N, Zeina CM, Gao X, Rees HA,
  Lin Z, et~al.
\newblock {Evolved Cas9 variants with broad PAM compatibility and high DNA
  specificity}.
\newblock \emph{Nature}, 556\penalty0 (7699):\penalty0 57--63, 2018.

\bibitem[Szczelkun et~al.(2014)Szczelkun, Tikhomirova, Sinkunas, Gasiunas,
  Karvelis, Pschera, Siksnys, and Seidel]{szczelkun2014}
Szczelkun MD, Tikhomirova MS, Sinkunas T, Gasiunas G, Karvelis T, Pschera P,
  Siksnys V, Seidel R.
\newblock {Direct observation of R-loop formation by single RNA-guided Cas9 and
  Cascade effector complexes}.
\newblock \emph{{Proc Natl Acad Sci USA}}, 111\penalty0 (27):\penalty0
  9798--9803, 2014.

\bibitem[Bradde et~al.(2020)Bradde, Nourmohammad, Goyal, and
  Balasubramanian]{bradde2020}
Bradde S, Nourmohammad A, Goyal S, Balasubramanian V.
\newblock The size of the immune repertoire of bacteria.
\newblock \emph{{Proc Natl Acad Sci USA}}, 117\penalty0 (10):\penalty0
  5144--5151, 2020.

\bibitem[Shvets and Kolomeisky(2017)]{shvets2017}
Shvets AA, Kolomeisky AB.
\newblock {Mechanism of genome interrogation: How CRISPR RNA-guided Cas9
  proteins locate specific targets on DNA}.
\newblock \emph{{Biophys J}}, 113\penalty0 (7):\penalty0 1416--1424, 2017.

\bibitem[Nishimasu et~al.(2015)Nishimasu, Cong, Yan, Ran, Zetsche, Li,
  Kurabayashi, Ishitani, Zhang, and Nureki]{nishimasu2015}
Nishimasu H, Cong L, Yan WX, Ran FA, Zetsche B, Li~Y, Kurabayashi A, Ishitani
  R, Zhang F, Nureki O.
\newblock {Crystal structure of \emph{Staphylococcus aureus} Cas9}.
\newblock \emph{{Cell}}, 162\penalty0 (5):\penalty0 1113--1126, 2015.

\bibitem[Jiang and Doudna(2015)]{jiang2015}
Jiang F, Doudna JA.
\newblock {The structural biology of CRISPR-Cas systems}.
\newblock \emph{Curr Opin Struct Biol}, 30:\penalty0 100--111, 2015.

\bibitem[Fineran and Charpentier(2012)]{fineran2012}
Fineran PC, Charpentier E.
\newblock {Memory of viral infections by CRISPR-Cas adaptive immune systems:
  acquisition of new information}.
\newblock \emph{Virology}, 434\penalty0 (2):\penalty0 202--209, 2012.

\bibitem[Bonomo(2020)]{bonomoPhD}
Bonomo ME.
\newblock \emph{{Investigating Modular Structure and Function in Biology: from
  Immunology to Cognition}}.
\newblock PhD thesis, Rice University, 2020.

\end{thebibliography}


\begin{thebibliography}{25}
\providecommand{\natexlab}[1]{#1}
\providecommand{\url}[1]{\texttt{#1}}
\expandafter\ifx\csname urlstyle\endcsname\relax
  \providecommand{\doi}[1]{doi: #1}\else
  \providecommand{\doi}{doi: \begingroup \urlstyle{rm}\Url}\fi

\bibitem[Farasat and Salis(2016)]{farasat2016}
Farasat I, Salis HM.
\newblock {A biophysical model of CRISPR/Cas9 activity for rational design of
  genome editing and gene regulation}.
\newblock \emph{PLoS Comp Biol}, 12\penalty0 (1):\penalty0 e1004724, 2016.

\bibitem[Bauer and Benham(1993)]{bauer1993}
Bauer WR, Benham CJ.
\newblock {The free energy, enthalpy and entropy of native and of partially
  denatured closed circular DNA}.
\newblock \emph{{J Mol Biol}}, 234\penalty0 (4):\penalty0 1184--1196, 1993.

\bibitem[Westra et~al.(2012)Westra, van Erp, K{\"u}nne, Wong, Staals, Seegers,
  Bollen, Jore, Semenova, Severinov, and et]{westra2012}
Westra ER, van Erp PB, K{\"u}nne T, Wong SP, Staals RH, Seegers CL, Bollen S,
  Jore MM, Semenova E, Severinov K, et~al.
\newblock {CRISPR immunity relies on the consecutive binding and degradation of
  negatively supercoiled invader DNA by Cascade and Cas3}.
\newblock \emph{{Mol Cell}}, 46\penalty0 (5):\penalty0 595--605, 2012.

\bibitem[Wang and Benham(2008)]{wang2008}
Wang H, Benham CJ.
\newblock Superhelical destabilization in regulatory regions of stress response
  genes.
\newblock \emph{PLoS Comp Biol}, 4\penalty0 (1):\penalty0 e17, 2008.

\bibitem[Wang(1979)]{wang1979}
Wang JC.
\newblock {Helical repeat of DNA in solution}.
\newblock \emph{{Proc Natl Acad Sci USA}}, 76\penalty0 (1):\penalty0 200--203,
  1979.

\bibitem[Rauzan et~al.(2013)Rauzan, McMichael, Cave, Sevcik, Ostrosky, Whitman,
  Stegemann, Sinclair, Serra, and Deckert]{rauzan2013}
Rauzan B, McMichael E, Cave R, Sevcik LR, Ostrosky K, Whitman E, Stegemann R,
  Sinclair AL, Serra MJ, Deckert AA.
\newblock {Kinetics and thermodynamics of DNA, RNA, and hybrid duplex
  formation}.
\newblock \emph{Biochemistry}, 52\penalty0 (5):\penalty0 765--772, 2013.

\bibitem[Leonard et~al.(1990)Leonard, Booth, and Brown]{leonard1990}
Leonard GA, Booth ED, Brown T.
\newblock {Structural and thermodynamic studies on the adenine. guanine
  mismatch in B-DNA}.
\newblock \emph{Nucleic Acids Res}, 18\penalty0 (19):\penalty0 5617--5623,
  1990.

\bibitem[Dickson et~al.(2000)Dickson, Burns, and Richardson]{dickson2000}
Dickson KS, Burns CM, Richardson JP.
\newblock {Determination of the free-energy change for repair of a DNA
  phosphodiester bond}.
\newblock \emph{J Biol Chem}, 275\penalty0 (21):\penalty0 15828--15831, 2000.

\bibitem[Ecevit et~al.(2010)Ecevit, Khan, and Goss]{ecevit2010}
Ecevit O, Khan MA, Goss DJ.
\newblock {Kinetic analysis of the interaction of b/HLH/Z transcription factors
  Myc, Max, and Mad with cognate DNA}.
\newblock \emph{Biochemistry}, 49\penalty0 (12):\penalty0 2627--2635, 2010.

\bibitem[Garcia-Viloca et~al.(2004)Garcia-Viloca, Gao, Karplus, and
  Truhlar]{garcia2004}
Garcia-Viloca M, Gao J, Karplus M, Truhlar DG.
\newblock How enzymes work: analysis by modern rate theory and computer
  simulations.
\newblock \emph{Science}, 303\penalty0 (5655):\penalty0 186--195, 2004.

\bibitem[Pollak and Talkner(2005)]{pollak2005}
Pollak E, Talkner P.
\newblock {Reaction rate theory: What it was, where is it today, and where is
  it going?}
\newblock \emph{Chaos}, 15\penalty0 (2):\penalty0 026116, 2005.

\bibitem[Chen et~al.(2008)Chen, Zhou, Qu, and Zhao]{chen2008}
Chen X, Zhou Y, Qu~P, Zhao XS.
\newblock {Base-by-base dynamics in DNA hybridization probed by fluorescence
  correlation spectroscopy}.
\newblock \emph{J Am Chem Soc}, 130\penalty0 (50):\penalty0 16947--16952, 2008.

\bibitem[Sanstead and Tokmakoff(2018)]{sanstead2018}
Sanstead PJ, Tokmakoff A.
\newblock {Direct observation of activated kinetics and downhill dynamics in
  DNA dehybridization}.
\newblock \emph{J Phys Chem B}, 122\penalty0 (12):\penalty0 3088--3100, 2018.

\bibitem[Gong et~al.(2018)Gong, Yu, Johnson, and Taylor]{gong2018}
Gong S, Yu~HH, Johnson KA, Taylor DW.
\newblock {DNA unwinding is the primary determinant of CRISPR-Cas9 activity}.
\newblock \emph{Cell Rep}, 22\penalty0 (2):\penalty0 359--371, 2018.

\bibitem[Szczelkun et~al.(2014)Szczelkun, Tikhomirova, Sinkunas, Gasiunas,
  Karvelis, Pschera, Siksnys, and Seidel]{szczelkun2014}
Szczelkun MD, Tikhomirova MS, Sinkunas T, Gasiunas G, Karvelis T, Pschera P,
  Siksnys V, Seidel R.
\newblock {Direct observation of R-loop formation by single RNA-guided Cas9 and
  Cascade effector complexes}.
\newblock \emph{{Proc Natl Acad Sci USA}}, 111\penalty0 (27):\penalty0
  9798--9803, 2014.

\bibitem[Nelson and Cox(2000)]{nelson2000}
Nelson D, Cox M.
\newblock \emph{{Bioenergetics and metabolism in Lehninger Principles of
  Biochemistry}}.
\newblock New York: Worth Publishers, 3rd edition, 2000.

\bibitem[Witz et~al.(2010)Witz, Dietler, and Stasiak]{witz2010}
Witz G, Dietler G, Stasiak A.
\newblock {Effect of DNA Supercoiling on DNA Decatenation and Unknotting
  Followed By Brownian Dynamics Simulations}.
\newblock \emph{Biophys J}, 98\penalty0 (3):\penalty0 62a, 2010.

\bibitem[Trun and Trempy(2009)]{trun2009}
Trun N, Trempy J.
\newblock \emph{{Fundamental bacterial genetics}}.
\newblock {John Wiley \& Sons}, 2009.

\bibitem[Slutsky and Mirny(2004)]{slutsky2004}
Slutsky M, Mirny LA.
\newblock Kinetics of protein-dna interaction: facilitated target location in
  sequence-dependent potential.
\newblock \emph{Biophys J}, 87\penalty0 (6):\penalty0 4021--4035, 2004.

\bibitem[Phillips et~al.(2012)Phillips, Kondev, Theriot, and
  Garcia]{phillips2012}
Phillips R, Kondev J, Theriot J, Garcia H.
\newblock \emph{{Physical biology of the cell}}.
\newblock Garland Science, 2012.

\bibitem[Stormo(2013)]{stormo2013}
Stormo GD.
\newblock {Modeling the specificity of protein-DNA interactions}.
\newblock \emph{{Quant Biol}}, 1\penalty0 (2):\penalty0 115--130, 2013.

\bibitem[Jones et~al.(2017)Jones, Leroy, Unoson, Fange, {\'C}uri{\'c}, Lawson,
  and Elf]{jones2017}
Jones DL, Leroy P, Unoson C, Fange D, {\'C}uri{\'c} V, Lawson MJ, Elf J.
\newblock {Kinetics of dCas9 target search in \emph{Escherichia} coli}.
\newblock \emph{{Science}}, 357\penalty0 (6358):\penalty0 1420--1424, 2017.

\bibitem[Singh et~al.(2016)Singh, Sternberg, Fei, Doudna, and Ha]{singh2016}
Singh D, Sternberg SH, Fei J, Doudna JA, Ha~T.
\newblock {Real-time observation of DNA recognition and rejection by the
  RNA-guided endonuclease Cas9}.
\newblock \emph{Nat Commun}, 7\penalty0 (1):\penalty0 1--8, 2016.

\bibitem[Sternberg et~al.(2014)Sternberg, Redding, Jinek, Greene, and
  Doudna]{sternberg2014}
Sternberg SH, Redding S, Jinek M, Greene EC, Doudna JA.
\newblock {DNA interrogation by the CRISPR RNA-guided endonuclease Cas9}.
\newblock \emph{{Nature}}, 507\penalty0 (7490):\penalty0 62, 2014.

\bibitem[Jiang et~al.(2013)Jiang, Bikard, Cox, Zhang, and
  Marraffini]{jiang2013}
Jiang W, Bikard D, Cox D, Zhang F, Marraffini LA.
\newblock {RNA-guided editing of bacterial genomes using CRISPR-Cas systems}.
\newblock \emph{Nat Biotechnol}, 31\penalty0 (3):\penalty0 233--239, 2013.

\end{thebibliography}

\end{document}


\singlespacing

\begin{center}

\noindent{\Large Supplemental Material}\\ 
\noindent {\bf\large A small PAM optimises target recognition in the \\ 
CRISPR-Cas immune system}

\vspace{3mm}
\noindent Melia E. Bonomo

\end{center}

\noindent{\bf Contents} 
\begin{table}[H]
\begin{tabular}{l l r}
\ref{sec:ModelDerivation}.  &  Model Derivation & ~~pg.~\pageref{sec:ModelDerivation} \\
 & \ref{sec:BFPeq}. Backwards Fokker-Planck Equation & pg.~\pageref{sec:BFPeq} \\
 & \ref{sec:pCleave}. Probability of CRISPR Cleavage & pg.~\pageref{sec:pCleave} \\
 & \ref{sec:tCleave}. Time to CRISPR Cleavage or Dissociation & pg.~\pageref{sec:tCleave} \\
\ref{sec:ModelParameters}.  & Model Parameters & pg.~\pageref{sec:ModelParameters} \\
\ref{sec:LandscapeResults}.  & Landscape Results & pg.~\pageref{sec:LandscapeResults} \\
\ref{sec:ConsecutiveMismatches}. & Consecutive Mismatches -- Extended Results & pg.~\pageref{sec:ConsecutiveMismatches} \\
\end{tabular}
\end{table}

\vspace{1em}

\section{Model Derivation}
\label{sec:ModelDerivation}

\subsection{Backwards Fokker-Planck Equation}
\label{sec:BFPeq}

We can imagine the CRISPR recognition reaction occurring on a one-dimensional landscape with forward rates $\lambda$ and backward rates $\mu$ to move between discrete states. State 0 represents the Cas9:crRNA unbound to target DNA, state 1 represents the bound state of the Cas9 to the PAM of the target DNA, states $i$ for $i=2...M-1$ represent bound states for each base pair of crRNA and target DNA, where $M-1$ is when crRNA and target DNA are fully bound, and state $M$ represents the post-cleavage state.

\begin{figure}[H]
\begin{center}
\includegraphics[width=0.5\textwidth]{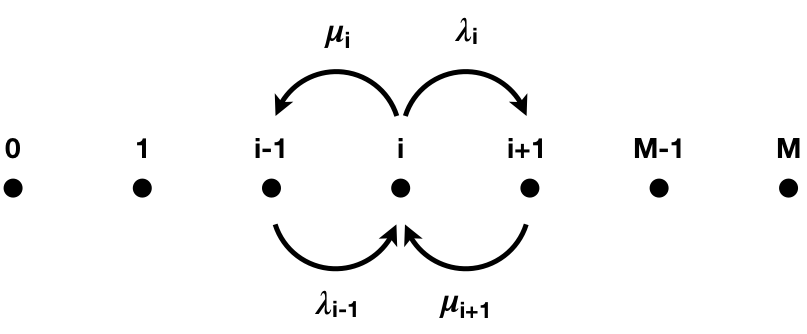}
\label{fig:lattice}
\end{center}
\end{figure}

Given the initial Cas9:PAM interaction at time $t=0$, we want to determine the probability $p_1$ and time $t_1$ to go from state 1 to cleavage in state $M$.  We can start with the one-dimensional Fokker-Planck equation for a stochastic process in operator form,
\begin{equation}
\dot{P}_i(t) = L P_i(t),
\label{eq:FFP} 
\end{equation}
where $P_i(t)$ is the probability of reaching $M$ from state $i$ at time $t$ and $L$ is the linear operator
\begin{equation*}
L_{i,j} = -\mu_i \delta_{i,j} - \lambda_i \delta_{i,j} + \mu_j \delta_{j,i+1} + \lambda_j \delta_{j,i-1}.
\end{equation*}
The probability of the CRISPR reaction being in state $M$ therefore changes over time according to
\begin{equation*}
\dot{P}_i(t) = -\mu_i P_{i} - \lambda_i P_{i} + \mu_{i+1} P_{i+1} + \lambda_{i-1} P_{i-1}.
\end{equation*}
Note that there is no drift term since the $\lambda_i$ and $\mu_i$ rates are different for each state $i$, and there is no diffusion term since we have discrete states.

Given an initial $P_i(0)$, we want to know the probability of being in state $M$ at a later time $t$, so we would integrate Eq.~\ref{eq:FFP} forward in time.  However, it will be simpler to consider a final $P_M(T)$ at $t=T$ and determine the probability of ending up in the target state from state $i$ at $t=0$ by integrating backwards in time.  The backwards Fokker-Planck equation is
\begin{equation}
-\dot{P}_i(t) = L^\intercal P_i(t),
\label{eq:BFP} 
\end{equation}
where $L^\intercal$ is the transpose of $L$,
\begin{equation*}
L_{i,j} = -\mu_i \delta_{i,j} - \lambda_i \delta_{i,j} + \mu_i \delta_{j+1,i} + \lambda_i \delta_{j-1,i},
\end{equation*}
which leads to
\begin{equation*}
- \dot{P}_i(t) = -\mu_i P_{i} - \lambda_i P_{i} + \mu_{i} P_{i-1} + \lambda_{i} P_{i+1}.
\end{equation*}

The boundary conditions of Eq.~\ref{eq:BFP} are
\begin{align*}
P_i(\infty) &= \delta_{i,M} \\
P_i(0) &= \text{probability to reach~} M \text{~from~} i \text{~by~} t_i<\infty,
\end{align*}
however it will be easier to integrate forward in time by shifting the boundaries to
\begin{align*}
P_i(0) &= \delta_{i,M} \\
P_i(-t) &=\text{probability to reach~} M \text{~from~} i \text{~by~} 0-(-t) = t_i.
\end{align*}
which changes the sign of $t$,
\begin{equation}
\dot{P}_i(t) = L^\intercal P_i(t).
\label{eq:BFPt} 
\end{equation}
We can now integrate Eq.~\ref{eq:BFPt} forward in time to obtain the First Passage Time, $T_i$, to reach state $M$ from state $i$,
\begin{equation*}
T_i = \int^{\infty}_0 t \frac{\partial P_i(t)}{\partial t}dt = t P_i(t)\Big|^{\infty}_0 - \int^{\infty}_0 P_i(t)dt,
\end{equation*}
however, this is undefined with
\begin{align*}
P_i(0) &= 0 \\
P_i(\infty) &= P^{\mathrm{eq}}_i,
\end{align*}
where $P^{\mathrm{eq}}_i$ is the equilibrium probability distribution for state $i$.

We can get the integral to converge by redefining the quantity that we want to calculate as the survival probability, $S_i(t)$, or the probability that the CRISPR reaction has not reached state $M$ at time $t$,
\begin{equation}
S_i(t) = P^{\mathrm{eq}}_i - P_i(t)
\label{eq:survival}
\end{equation}
\begin{align*}
S_i(0) &= P^{\mathrm{eq}}_i \\
S_i(\infty) &= 0.
\end{align*}
The First Passage Time is now calculated as
\begin{equation}
T_i = - \int^{\infty}_0 t \frac{\partial S_i(t)}{\partial t}dt = - t S_i(t)\Big|^{\infty}_0 + \int^{\infty}_0 S_i(t)dt = \int^{\infty}_0 S_i(t)dt.
\label{eq:fpt}
\end{equation}

We can substitute Eq.~\ref{eq:survival} and its derivative into Eq.~\ref{eq:BFPt} to obtain
\begin{align*}
-\dot{S}_i(t) &= L^\intercal (P^{\mathrm{eq}}_i - S_i(t)) \\
&=  -\mu_i P^{\mathrm{eq}}_i - \lambda_i P^{\mathrm{eq}}_i + \mu_{i} P^{\mathrm{eq}}_{i-1} + \lambda_{i} P^{\mathrm{eq}}_{i+1} + \mu_i S_{i} + \lambda_i S_{i} - \mu_{i} S_{i-1} - \lambda_{i} S_{i+1}.
\end{align*}
Since the equilibrium probability distribution of reaching state $M$ from state $i$ is
\begin{equation*}
P^{\mathrm{eq}}_i = \mu_i P^{\mathrm{eq}}_{i-1} +(1-\mu_i-\lambda_i)P^{\mathrm{eq}}_i + \lambda_{i} P^{\mathrm{eq}}_{i+1}
\end{equation*}
\begin{equation}
0 = \mu_{i} P^{\mathrm{eq}}_{i-1} - \mu_i P^{\mathrm{eq}}_i - \lambda_i P^{\mathrm{eq}}_i + \lambda_{i} P^{\mathrm{eq}}_{i+1},
\label{eq:P}
\end{equation}
we can determine that
\begin{align*}
\dot{S}_i(t) &= - \mu_i S_{i} - \lambda_i S_{i} + \mu_{i} S_{i-1} + \lambda_{i} S_{i+1} \\
\dot{S}_i(t) &= L^\intercal S_i(t).
\end{align*}
Now Eq.~\ref{eq:fpt} can be solved as
\begin{align}
T_i &= \int^{\infty}_0 S_i(t)dt = \int^{\infty}_0 \frac{\dot{S}_i(t)}{L^\intercal}dt = \frac{S_i(t)}{L^\intercal}\Big|^{\infty}_0 \nonumber \\
S_i(\infty) - S_i(0) &= L^\intercal T_i \nonumber \\
- S_i(0) &= - \mu_i T_{i} - \lambda_i T_{i} + \mu_{i} T_{i-1} + \lambda_{i} T_{i+1} \nonumber \\
P^{\mathrm{eq}}_{i} &= \mu_i T_{i} + \lambda_i T_{i} - \mu_{i} T_{i-1} - \lambda_{i} T_{i+1}. \label{eq:time}
\end{align}
The $T_i$ in Eq.~\ref{eq:time} can be divided by the conditional probability $P^{\mathrm{eq}}_{i}$ for reaching state $M$ ($p_i$), state 0 ($1-p_i$), or either state ($1-p_i+p_i=1$) to obtain the $t_i$ that we want.

\subsection{Probability of CRISPR Cleavage}
\label{sec:pCleave}

Given Eq.~\ref{eq:P} and the fact that states 0 and $M$ are considered absorbing states, the probability $p_i$ of reaching state $M$ from $i$ is
\begin{align*}
p_0 &= 0 \\
p_M &= 1
\end{align*}
\begin{equation}
0 = \mu_{i} p_{i-1} - \mu_i p_i - \lambda_i p_i + \lambda_{i} p_{i+1}.
\label{eq:p_i}
\end{equation}
As Eq.~\ref{eq:p_i} is a recursive equation, we can extract a concise equation for our desired probability. For i = 1,...,M let
\begin{equation}
z_i = p_i - p_{i-1},
\label{eq:z_i}
\end{equation}
and it follows that
\begin{equation*}
\sum^M_{i=1} z_i = p_1 - p_0 + p_2 - p_1 + ... + p_M - p_{M-1} = p_M - p_0 = 1.
\end{equation*}
Using Eq.~\ref{eq:p_i} and simplifying algebraically, we find that $z_{i+1} = \gamma_i z_i$ where $\gamma_i = \mu_i / \lambda_i$. With this relationship and Eq.~\ref{eq:z_i}, we obtain
\begin{align*}
z_1 &= p_1 \\
z_2 &= \gamma_1 z_1 = \gamma_1 p_1 \\
z_3 &= \gamma_2 z_2 = \gamma_2 \gamma_1 p_1,
\end{align*}
and so on. The sum of all $z$ values is then
\begin{align}
1 = \sum^{M}_{i = 1} z_i &= p_1 + \gamma_1 p_1 + \gamma_2 \gamma_1 p_1 + ... \nonumber \\
1 &= p_1(1 + \gamma_1 + \gamma_2 \gamma_1 + ...) \nonumber \\
p_1 &= \frac{1}{1 + \gamma_1 + \gamma_2 \gamma_1 + ...}  \nonumber \\
p_1 &= \frac{1}{1 + \sum\limits^{M-1}_{i=1} \prod\limits^{i}_{j=1} \gamma_j}, \label{eq:p_1}
\end{align}
where Eq.~\ref{eq:p_1} is the probability that, given PAM binding in state $i=1$, all crRNA nucleotides will be bound to those of the target DNA, and target DNA will be cleaved in state $M$.  The equation for this probability from any state $i$ is then
\begin{equation}
p_i = p_1 \Big( 1 + \sum^{i-1}_{j=1} \prod^{j}_{k=1} \gamma_k \Big).
\end{equation}

\subsection{Time to CRISPR Cleavage or Dissociation}
\label{sec:tCleave}

Given Eq.~\ref{eq:time} and the fact that states 0 and $M$ are considered absorbing states, the time $T_i$ to reach state $M$ from $i$ is
\begin{align*}
T_0 &= \infty \\
T_M &= 0
\end{align*}
\begin{align}
p_{i} &= \mu_i T_{i} + \lambda_i T_{i} - \mu_{i} T_{i-1} - \lambda_{i} T_{i+1} \label{eq:beforeT_i} \\
T_i &= \frac{p_i}{\mu_i + \lambda_i} + \frac{\lambda_i}{\mu_i + \lambda_i}T_{i+1} + \frac{\mu_i}{\mu_i + \lambda_i}T_{i-1}  \label{eq:T_i}
\end{align}
As Eq.~\ref{eq:T_i} is a recursive equation, we can extract a concise equation for our desired time. For i = 1,...,M-1 let
\begin{equation}
y_i = T_i - T_{i+1},
\label{eq:y_i}
\end{equation}
and it follows that
\begin{equation*}
\sum^{M-1}_{i=1} y_i = T_1 - T_2 + T_2 - T_3 + ... + T_{M-1} - T_{M} = T_1 - T_M = T_1.
\end{equation*}
Using Eq.~\ref{eq:T_i} and simplifying algebraically, we find that $y_{i} = \frac{p_i}{\lambda_i} + \gamma_i y_{i-1}$. With this relationship and Eq.~\ref{eq:y_i}, we obtain
\begin{align*}
y_1 &= \frac{p_1}{\lambda_1} - \gamma_1 T_{1} \\
y_2 &= \frac{p_2}{\lambda_2} + \gamma_2 y_{1} =  \frac{p_2}{\lambda_2} + \gamma_2 \frac{p_1}{\lambda_1} - \gamma_2 \gamma_1 T_1\\
y_3 &= \frac{p_3}{\lambda_3} + \gamma_3 y_{2} =  \frac{p_3}{\lambda_3} + \gamma_3 \frac{p_2}{\lambda_2} + \gamma_3 \gamma_2 \frac{p_1}{\lambda_1} - \gamma_3 \gamma_2 \gamma_1 T_1,
\end{align*}
and so on. The sum of all $y$ values is then
\begin{align}
T_1 = \sum^{M-1}_{i = 1} y_i &= \frac{p_1}{\lambda_1} - \gamma_1 T_{1} + \frac{p_2}{\lambda_2} + \gamma_2 \frac{p_1}{\lambda_1} - \gamma_2 \gamma_1 T_1 + \frac{p_3}{\lambda_3} + \gamma_3 \frac{p_2}{\lambda_2} + \gamma_3 \gamma_2 \frac{p_1}{\lambda_1} - \gamma_3 \gamma_2 \gamma_1 T_1 + ... \nonumber \\
T_1 &= \frac{(\frac{p_1}{\lambda_1} + \frac{p_2}{\lambda_2} + \frac{p_3}{\lambda_3} + ...) + [\frac{p_1}{\lambda_1}(\gamma_2 + \gamma_3 \gamma_2 + ...) + \frac{p_2}{\lambda_2}(\gamma_3 + \gamma_4 \gamma_3 + ...) + ...]}{1 + \gamma_1 + \gamma_2 \gamma_1 + \gamma_3 \gamma_2 \gamma_1...} \nonumber \\
T_1 &= \frac{\sum\limits^{M-1}_{i=1}\frac{p_i}{\lambda_i} + \sum\limits^{M-2}_{i=1}\frac{p_i}{\lambda_i} \Bigg( \sum\limits^{M-1}_{j=i+1} \prod\limits^{j}_{k=i+1} \gamma_k  \Bigg) }{1 + \sum\limits^{M-1}_{i=1} \prod\limits^{i}_{j=1} \gamma_j}. \label{eq:T_1}
\end{align}
Dividing Eq.~\ref{eq:T_1} by the probability $p_1$ of reaching state $M$ from state $1$, yields
\begin{equation}
t_{(1,M)} = \sum\limits^{M-1}_{i=1}\frac{p_i}{\lambda_i} + \sum\limits^{M-2}_{i=1}\frac{p_i}{\lambda_i} \Bigg( \sum\limits^{M-1}_{j=i+1} \prod\limits^{j}_{k=i+1} \gamma_k  \Bigg),
\label{eq:t_1toM}
\end{equation}
which is the time from the PAM binding in state $i=1$ to target DNA cleavage in state $M$.

Conversely, if we consider $1-p_i$, which is the probability of reaching state $0$ from state $i$ in Eq.~\ref{eq:beforeT_i}, and we
divide the resulting equivalent of Eq.~\ref{eq:T_1} by $1-p_1$, the probability of reaching state $0$ from state $1$, we obtain
\begin{equation}
t_{(1,0)} = \Bigg[ \sum\limits^{M-1}_{i=1}\frac{1-p_i}{\lambda_i} + \sum\limits^{M-2}_{i=1}\frac{1-p_i}{\lambda_i} \Bigg( \sum\limits^{M-1}_{j=i+1} \prod\limits^{j}_{k=i+1} \gamma_k  \Bigg) \Bigg] \frac{p_1}{1-p_1},
\label{eq:t_1to0}
\end{equation}
which is the time from the PAM binding in state $i=1$ to dissociation from the PAM in state $i=0$.

\section{\bf Model Parameters}
\label{sec:ModelParameters}

The free energy and reaction parameters utilised in the model are summarised in Table~\ref{tab:parameters}. 

\begin{table}[H]
\caption{Energy and kinetic parameters used in the CRISPR model and the experimental source from which they were obtained or estimated. The dsDNA and RNA:DNA melting and binding energies are simplified to be the same for A:T and C:G base pairs (bp). Note that the forward and reverse attempt rates $A$ for interrogating the dsDNA protospacer ($\lambda_i$ for $i=1...M-2$ and $\mu_i$ for $i=2...M-1$) were assumed to be equivalent. $k_B T = 0.62$ kcal/mol is used.}
\vspace{1em}
\scriptsize
\begin{tabular}{llll}
\hline
{\bf Free Energy}                         & {\bf Description}                                                                   & {\bf Value}                                & {\bf Source}                                         \\
\hline
$\Delta{}G_{\text{MeltMatch}}$     & melting 1 dsDNA bp                                                             & 3 kcal/mol & \cite{farasat2016}            \\
$\Delta{}G_{\text{Supercoiling}}$  & topological constraint per bp                                     & -0.8 kcal/mol                        & \cite{bauer1993,westra2012,wang2008,wang1979}              \\
$\Delta{}G_{\text{BindMatch}}$     & binding 1 RNA:DNA bp                                                            & $-\Delta{}G_{\text{MeltMatch}}$      & \cite{rauzan2013}             \\
$\Delta{}G_{\text{BindMismatch}}$  & binding 1 mismatched RNA:DNA bp				 & $- 1/2~\Delta{}G_{\text{MeltMatch}}$ & \cite{rauzan2013,leonard1990} \\
$\Delta{}G_{\text{Cleavage}}$ & break DNA phosphodiester bonds                                 & 6 kcal/mol per strand                & \cite{dickson2000}           \\
 \hline
  & & & \\
 \hline
 {\bf Rate}                                                      & {\bf Attempt Rate} & {\bf Barrier Height} & {\bf Source}              \\
 \hline
PAM dissociation ($\mu_{1}$)                            & $A_{1,0}$ = 6x$10^{12}$            & $\Delta{}E_{0,1}$ = 3.4 kcal/mol                      & \cite{ecevit2010,garcia2004,pollak2005}            \\
dsDNA separation initiation ($\lambda_{1}$)	 & $A_{1,2} = A_{2,1}$               & $\Delta{}E_{1,2}$ = 10.2 kcal/mol                     & \cite{bauer1993} \\
dsDNA melting ($\lambda_{i}$, for $i=2...M-2$)	 & $A_{i,i+1}$ = $A_{i,i-1}$ = 1x$10^9$               & $\Delta{}E_{i,i+1}$ = 0.9 kcal/mol                     & \cite{chen2008,sanstead2018} \\
dsDNA cleavage ($\lambda_{M-1}$)                            & $A_{M-1,M}$ = 1x$10^9$               & $\Delta{}E_{M-1,M}$ = 12 kcal/mol                       & \cite{gong2018}  \\
\hline
\end{tabular}
\label{tab:parameters}
\end{table}

There is a reduced free energy for Cas9 to separate a double-stranded DNA (dsDNA) target that is associated with supercoiling~\cite{szczelkun2014,farasat2016,westra2012},
\begin{equation}
\Delta{}G_{\text{dsDNAseparation}} = \Delta{}G_{\text{MeltMatch}} + \Delta{}G_{\text{Supercoiling}},
\end{equation}
where $\Delta{}G_{\text{MeltMatch}}$ is the free energy associated with melting individual base pairs of relaxed dsDNA. It is energetically favourable to separate DNA strands that are topologically constrained due to negative supercoiling~\cite{farasat2016}.  For easy strand separation and compaction, cellular DNA is generally kept 5\% to 7\% under-wound, resulting in negative superhelical twists~\cite{nelson2000}.  In fact, positive supercoiling helps to protect extreme thermophiles from spontaneous thermal denaturation~\cite{witz2010}.  During lysogenic infection, the phage injects its DNA into the cell, the DNA joins its ends and circularises, and the circular DNA then becomes negatively supercoiled by means of the host's machinery in order to more easily integrate itself into the bacterial genome~\cite{trun2009}.  Though \emph{in vivo} supercoiling is a dynamic quantity~\cite{westra2012}, we estimate an average free energy of supercoiling for $n$ DNA base pairs to be
\begin{equation}
\Delta{}G_{\text{Supercoiling}} = n \frac{qRT}{h^2_0} \sigma,
\end{equation}
where $R$ is the gas constant and $T=37^{\circ}$C is the temperature inside the cell~\cite{bauer1993,westra2012}. The average superhelical density $\sigma$ is taken to be -0.06~\cite{wang2008}, the parameter $q$ is an experimentally determined coefficient $\approx 1000$~\cite{bauer1993}, and the DNA helix repeat $h_0$ is 10.4 bp per turn~\cite{wang1979}.

In general for protein:DNA interaction energetics, the change in free energy related to binding has specific $E(\overrightarrow{b})$ and nonspecific $E_{\text{ns}}$ components, 
\begin{equation}
\Delta{}G_{\text{binding}}(N) = E(\overrightarrow{b}) + E_{\text{ns}}, 
\end{equation}
where $\overrightarrow{b}$ is a DNA sequence $b_i, ... b_{i+N-1}$ of length $N$~\cite{slutsky2004}.  The nonspecific binding energy does not depend on the actual nucleotide sequence, but rather accounts for the interaction between the protein and the DNA's phosphate backbone.  The specific binding energy is linearly related to the individual contribution of the protein interacting with each nucleotide, in which the energy change with respect to binding the matching target is
\begin{equation}
E(\overrightarrow{b}) - E(\overrightarrow{b})_{\text{Match}} = \sum^{N}_{i=1} \epsilon(i,b_i),
\end{equation}
where $\epsilon(i,b_i)$ is the energy penalty of mismatching base $b_i$ in position $i$~\cite{slutsky2004,phillips2012}.  The value of $\epsilon(i,b_i)$ is obtained according to the protein's position weight matrix, which can be defined in several ways~\cite{stormo2013}.   This matrix approximates protein specificity by assigning a score to each base at each position, relative to the matching target sequence.  

At present, a position weight matrix has not been experimentally determined for Cas9, so we proceed as follows.  For PAM binding $\Delta{}G_{\text{PAM}}$, the scores are defined as position- and base-independent match rewards that each contribute to the binding by an amount of energy $\epsilon$ such that
\begin{equation}
\Delta{}G_{\text{PAM}}(n_{\text{match}}) = \epsilon * n_{\text{match}} + E_{\text{ns}},
\label{eq:PAM}
\end{equation}
where $n_{\text{match}}$ is the number of matching nucleotides between the Cas9's DNA-interacting domain and the DNA target. Dissociation rates for Cas9 from matching and non-matching NGG PAMs were obtained from published CRISPR experiments~\cite{jones2017,singh2016}. A generic barrier height and attempt rate for protein:DNA dissociation were estimated from \cite{ecevit2010} and transition state theory for protein unfolding~\cite{pollak2005,garcia2004}. The Arrhenius relation was then used to calculate $\Delta{}G_{\text{PAM}}$ for binding matching and non-matching 3nt-PAMs, and these values were used to obtain $\epsilon$ and $E_{\text{ns}}$ from a linear fit of Eq.~\ref{eq:PAM},
\begin{equation}
\Delta{}G_{\text{PAM}}(n_{\text{match}}) = -1.63~\text{kcal/mol} * n_{\text{match}} -14.7~\text{kcal/mol}.
\label{eq:PAMfit}
\end{equation}
Given experimental observations of negligible Cas9 interaction with the DNA sequence in the absence of a PAM~\cite{jones2017,sternberg2014}, $\Delta{}G_{\text{PAM}}(0)$ was set to 0. 
It is important to note that this fitting method and use of Eq.~\ref{eq:PAMfit} to define the free energy change greatly simplify Cas9:PAM binding. Endogenous Cas9 PAMs, such as the NGG of \emph{Streptococcus pyogenes} or the NGRRT of \emph{Staphylococcus aureus}, have complex position- and base-dependent energy penalties for mismatches that impact binding~\cite{jiang2013,farasat2016,sternberg2014}. However, the objective of this model is to investigate how the probability of cleaving a DNA sequence, and the time for cleaving or dissociating from that sequence, depends on the amount of initial energy associated with the first module (i.e., the bound Cas9:PAM). Eq.~\ref{eq:PAMfit} allows us to probe how the probability and times are impacted when that first module's energy is increased and decreased in approximate units of mismatches.

\section{Landscape Results}
\label{sec:LandscapeResults}

\begin{figure}[H]
{\bf A} \includegraphics[width=0.88\textwidth]{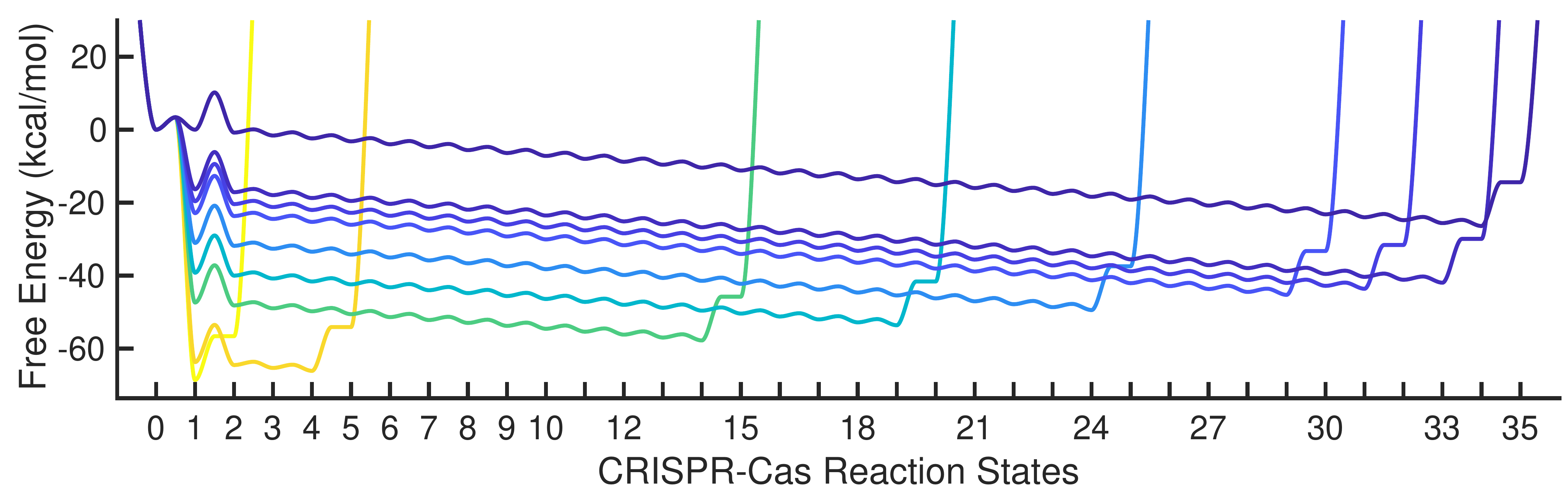}
\includegraphics[width=0.08\textwidth]{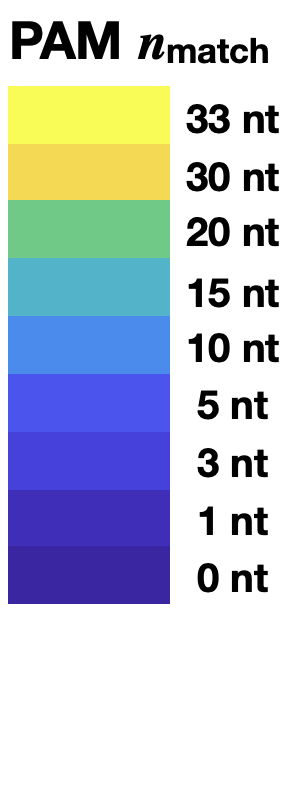} \\
{\bf B} \includegraphics[width=0.88\textwidth]{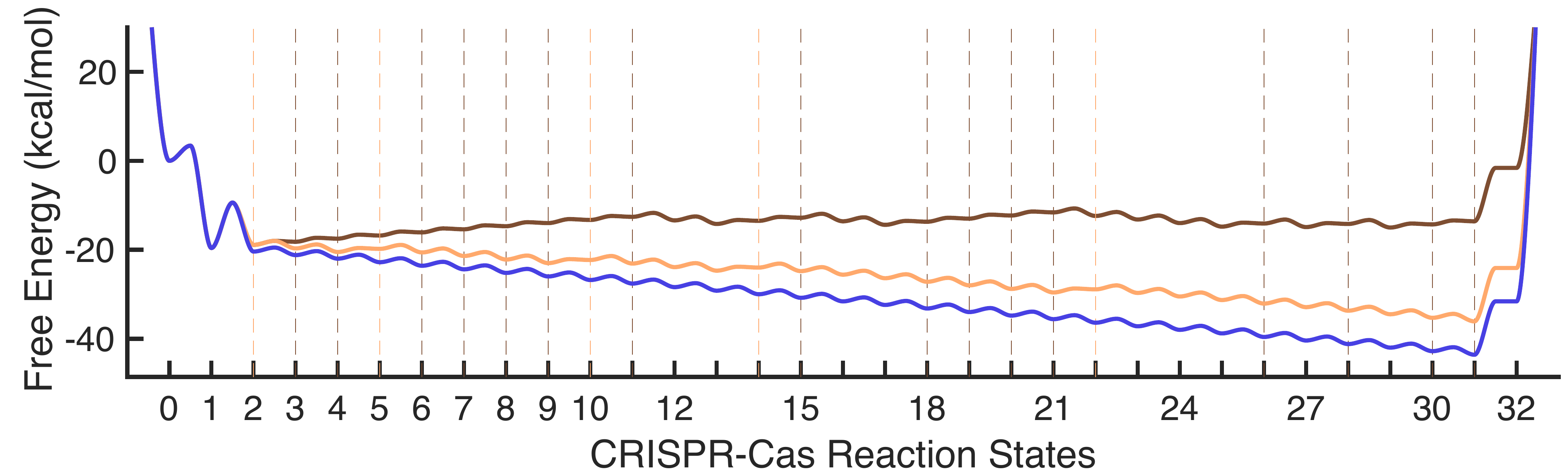}
\caption{{\bf (A)} Reaction landscapes calculated for nine representative CRISPR systems that had PAMs of different lengths. In all cases, the target DNA sequence's PAM and protospacer were perfect matches to the Cas9:crRNA. 
{\bf (B)} Reaction landscapes calculated for the 3nt-PAM CRISPR when faced with a target DNA sequence that had no (blue), 5 (tan), and 20 (brown) mismatches. The dotted lines designate where the mismatches were located (coloured to match their corresponding landscape).}
\label{fig:landscapes}
\end{figure}

The reaction landscapes for the CRISPR interrogation of target DNA were defined by the parameters in Table~\ref{tab:parameters}, where $\Delta{}G_i$ was the depth of each state $i$,
\begin{align}
\Delta{}G_1 &= \Delta{}G_{\textrm{PAM}}(n_{\text{match}}), \\
\Delta{}G_{1 > i > M} &= \Delta{}G_{\textrm{dsDNAseparation}} + \Delta{}G_{\textrm{BindDNA}}, \\
\Delta{}G_{M} &= \Delta{}G_{\textrm{Cleavage}},
\end{align}
the forward barriers were $\Delta{}E_{i-1,i}$, and the reverse barriers were calculated as
\begin{equation}
\Delta{}E_{i,i-1} = \Delta{}E_{i-1,i} - \Delta{}G_i,
\end{equation}
as stated in the main text. For a target DNA sequence that was a perfect match, the landscapes all trended downwards (Figure~\ref{fig:landscapes}A). Mismatches caused the landscapes to trend upwards, and the more mismatches there were, the more energetically unfavourable it was for the CRISPR to cleave the DNA sequence (Figure~\ref{fig:landscapes}B). It has been suggested that near-perfect complementarity between the crRNA and target DNA in the PAM-proximal region lowers the energy needed to continue binding the rest of the target, such that it is less than that of the reverse unzipping reaction~\cite{sternberg2014}.

\section{Consecutive Mismatches -- Extended Results}
\label{sec:ConsecutiveMismatches}

\begin{figure}[H]
\begin{center}
{\bf A}\includegraphics[width=0.99\textwidth]{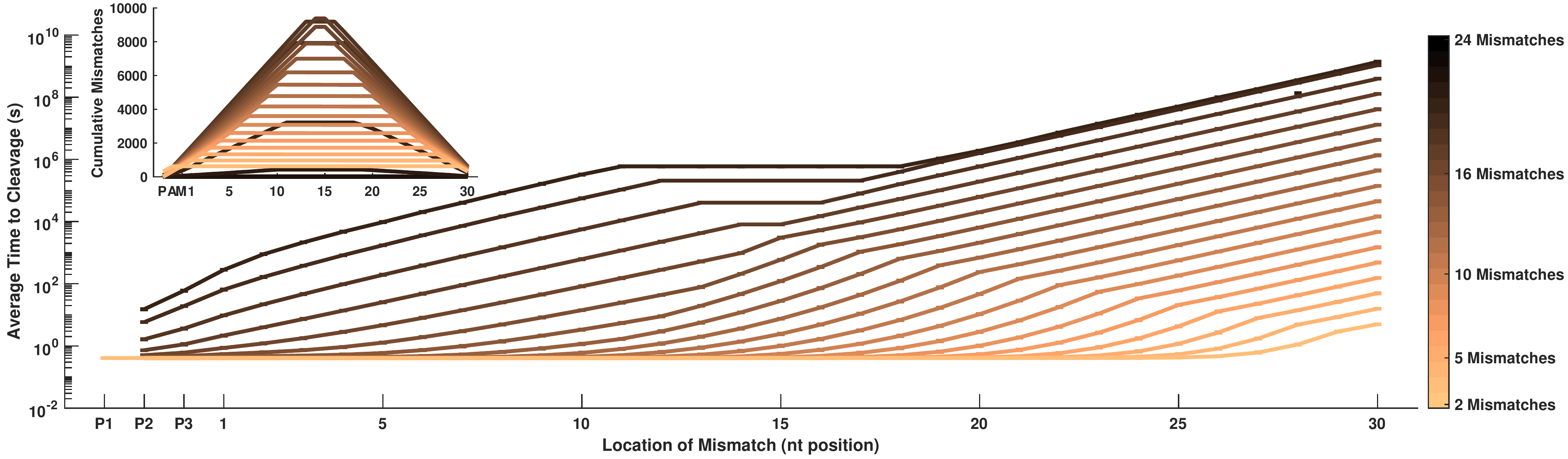} \\
\vspace{3mm}
{\bf B}\includegraphics[width=0.99\textwidth]{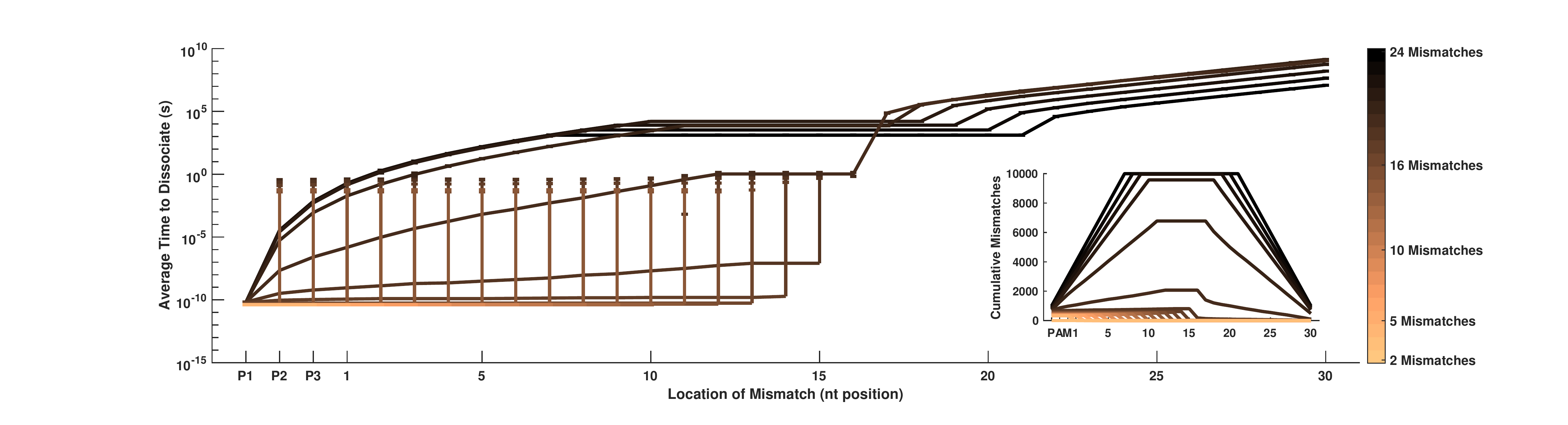}
\caption{Mismatch tolerance for the 3-nt PAM system illustrated by averaging over {\bf (A)} cleavage time or {\bf (B)} dissociation time when there were mismatches at each sequence location. Each curve is a geometric mean from 10,000 iterations of testing the specified number of consecutive mismatches (from 2 to 24 mismatches) along all possible positions in the sequence, and error bars represent standard error. The insets for each figure show the cumulative number of mismatches at each sequence location when {\bf (A)} cleavage occurred and {\bf (B)} dissociation occurred. (In other words, if each curve of the insets was divided by the total number of cumulative mismatches tested at each sequence location, the inset in {\bf (A)} would show the cleavage frequency plotted in Fig.~3 of the main text, and the inset in {\bf (B)} would show the dissociation frequency.) P$i$ denotes the location of PAM nucleotide $i$, and the $x$-axis restarts at 1 for nucleotides after the PAM.}
\label{fig:mismatches3ntPAM}
\end{center}
\end{figure}

\bibliographystyle{unsrtnatabbv}
\bibliography{Bonomo_CRISPR}